# AC vs. DC Electrophoretic Deposition of Hydroxyapatite on Titanium

V Ozhukil Kollath[a,b], Q Chen[c], R Closset[b], J Luyten[a,1], K Traina[b,2], S Mullens[a,*], A R Boccaccini[c], R Cloots[b]

[a] Sustainable Materials Management, Flemish Institute for Technological Research (VITO), 2400 Mol, Belgium; [b] GREEnMat, Department of Chemistry, University of Liège, 4000 Liège, Belgium; [c] Institute of Biomaterials, Department of Materials Science and Engineering, University of Erlangen-Nuremberg, 91058 Erlangen, Germany.

*Corresponding author. Email – steven.mullens@vito.be; Tel.: 0032 14335668; Fax: 0032 14321186

***Abstract***

*The bio-inertness of titanium and its alloys attracts their use as bone implants. However a bioactive coating is usually necessary for improving the bone bonding of such implants. In this study, electrophoretic deposition (EPD) of hydroxyapatite (HA) powder on titanium plate was performed using butanol as solvent under direct current (DC) and alternating current (AC) fields. The zeta potential of the suspensions was measured to understand their stability and the charge on the particles. Coating thickness was varied by adjusting the voltage and time of deposition. Surface morphology and cross section thickness were studied using scanning electron microscopy and image analysis software. Surface crack density was calculated from the micrographs. The results showed that the samples of similar thickness have higher grain density when coated using AC as compared to DC EPD. This facile but novel test proves the capability of AC-EPD to attain denser and uniform HA coatings from non-aqueous medium.*

## 1. Introduction

Electrophoretic deposition (EPD) is widely used as an easy and cost effective technique for a variety of coating applications.[1-5] One of the interesting applications include surface modification of metallic bone implant materials. Coating of these implants using biocompatible calcium phosphate (CaP) simultaneously improves the bioactivity and reduces the potentially harmful metal ions release.[6-8] Other industrially applied CaP coating techniques include plasma spraying deposition, electrostatic spray deposition and pulsed laser deposition.[9] However, EPD has some advantages over reported techniques as it operates at ambient temperature, is comparatively cheap and has a potential to coat the interior of porous materials (e.g. tissue scaffolds).

Direct current-EPD (DC-EPD) and alternating current-EPD (AC-EPD) are methods in which the applied voltage is supplied from a DC field or an AC field, respectively. For DC-EPD technique, the deposition of charged particles occurs in suspension to the oppositely charged electrode, under the influence of constant electric field. In AC-EPD technique, the direction of electric field is reversed periodically.[10-12] This accounts for oscillation and migration of powder particles in the suspension between electrodes. This oscillating migration is dependent on the frequency and asymmetry of the wave applied.[13] Stability of suspension and charge of the particles dispersed are two important criteria for EPD in general. These are mutually related since the charge on the particles increases the stability of suspensions due to electrostatic interactions.

This paper illustrates a comparative study between the AC-EPD and DC-EPD of hydroxyapatite (HA) on titanium (Ti) plates with butanol as the suspending medium. HA is a well accepted biocompatible phase of CaP. Water is not preferred as a medium of dispersion for HA, as the stability of the suspension is poor without the help of dispersants and results in immediate sedimentation. Butanol on the other hand provides sufficient stability for HA suspension during the EPD process. Butanol was preferred over ethanol in order to lower the evaporation rate which subsequently reduces cracking during drying of HA deposits. Deposited layers were then sintered in order to obtain sufficient adhesion with the substrate. They were further characterized using X-ray diffraction (XRD) analysis and scanning electron microscopy (SEM). Micrographs are analyzed to estimate the surface crack density and deposition density (based on the pore area percentage of the cross-section). The possible coating mechanism for HA in both techniques has been discussed from the morphology of surface and cross-section.

## 2. Materials and Methods

Commercially available Ti plates (thickness = 0.6 mm) were cut to 10 X 30 mm substrate and used as both working and counter electrodes. These substrates were cleaned with ethanol, acetone and high purity water in an ultrasonic bath (25 KHz, 100 % sweep; Elma, Fischer Bioblock Scientific) for 5 min each. After the cleaning process, the plates were immersed in pure butanol until further use. A 5 % (wt/wt) suspension was prepared from commercially available HA powder (Merck, Complexometric assay > 90 %) mixed with extra pure butanol (≥ 99 %). The suspension was ultrasonicated (40 sec at 75 % amplitude using 13 mm ultrasonic horn, 400 W power; Hielscher UP400S) prior to deposition for homogenization and stirred using a magnet at 200 rpm until use. The distance between the electrodes was kept constant at 10 mm. AC signal used was square type with an asymmetry factor 4, attained through a programmable function generator (HP 3314A). The frequency of the signal was 1000 Hz. A description of samples and the deposition parameters used are given in table 1.

Table 1 Voltages and times of deposition used in this study and the corresponding samples names

| Method | Voltage (V) | Time (sec) | Sample name |
|---|---|---|---|
| DC-EPD | 220 | 10 | DC-EPD-1 |
| | 260 | 10 | DC-EPD-2 |
| | 300 | 10 | DC-EPD-3 |
| | 340 | 10 | DC-EPD-4 |
| AC-EPD | 100 | 30 | AC-EPD-1 |

| | | |
|---|---|---|
| 150 | 30 | AC-EPD-2 |
| 100 | 60 | AC-EPD-3 |

Zeta potential measurements were carried out using zeta probe DT1200 (Dispersion Technologies Inc.) with 5 % (wt/wt) suspensions of hydroxyapatite (HA) prepared in pure butanol and kept under stirring conditions. Calculated zeta potential is the average of 15 values measured at 3 different depths within the suspension. Prior to zeta potential measurement, the suspension was ultrasonicated as described above. Coated substrates were sintered (Gero Hochtemperaturöfen GmbH, Germany) at 900 °C for 2 h in high purity argon atmosphere. The heating cycle included a gradual rise in temperature at a rate of 1 °C $min^{-1}$ until 200 °C (10 min) followed by 5 °C $min^{-1}$ until 900 °C. The lower heating rate up to 200 °C was used to minimize the solvent evaporation and hence reduce the cracking. Cooling rate was fixed at 5 °C $min^{-1}$. This sintering temperature was selected from various observations reported in literature.[14-16] Sintering temperature for full densification of commercial HA (>1000 °C) was compromised for reducing the ion migration from Ti substrate which leads to the decomposition of HA phase. This study focuses mainly on the deposition methods to assess the effect of applied fields (AC or DC) on the density and cracking behavior of deposited layer.

Selected samples were surface characterized using X-ray diffractometer (X'pert Pro, PANalytical) after sintering and compared with the spectrum of substrate plate. Cross-sections of samples were prepared for coating thickness measurements, after embedding these samples in epoxy resin. Surface morphology and cross-sections of the coated samples were procured using field emission scanning electron microscope (FESEM; JSM-6340F, Jeol). Thickness of the coating was calculated using image processing software Gwyddion (version 2.28), as an average of 30 measurements from 3 different areas captured. The percentage area of surface cracks and percentage pore area along the thickness of the coatings were calculated using ImageJ software (version 1.46). For crack area percentage, a Gaussian blur was applied before applying an automatic MaxEntropy threshold[17] and was calculated from the ratio between the black pixels and the total pixels in the image. The results calculated from three micrographs were averaged. For percentage pore area calculation using ImageJ, a rectangular area was selected on the HA region of selected micrographs. Threshold value was adjusted to best fit during particle area calculation. The pores are considered as particles here and the resulting area percentage was calculated from the area of pores detected within the selection.

# 3. Results and Discussion

The zeta potential of HA-Butanol suspension was 57.4 ± 0.4 mV (mean ± standard deviation). The suspension showed no signs of particle sedimentation (visual observation) for at least 24 h. The XRD pattern (figure 1) of sintered coating revealed HA as the major crystalline phase and tricalcium phosphate (TCP) as a minor phase. This decomposition of HA during sintering is reported to be caused by the partial dehydration mechanism or by metal ion exchange at the interface during the sintering temperature.[14, 16, 18-20] Rutile phase ($TiO_2$) was also observed in the sintered samples. However, no major peaks of rutile phase were observed in the XRD spectrum of substrate plate (figure A.1). Since the sintering was performed under Ar atmosphere, this eliminates the possibility of surface oxidation due to sintering atmosphere. However the TCP phase observed in figure 1 possibly indicates that the ion migration during sintering facilitates the formation of rutile phase at the HA-Ti interface. This is in concordance with the findings of Wei et al.[15]

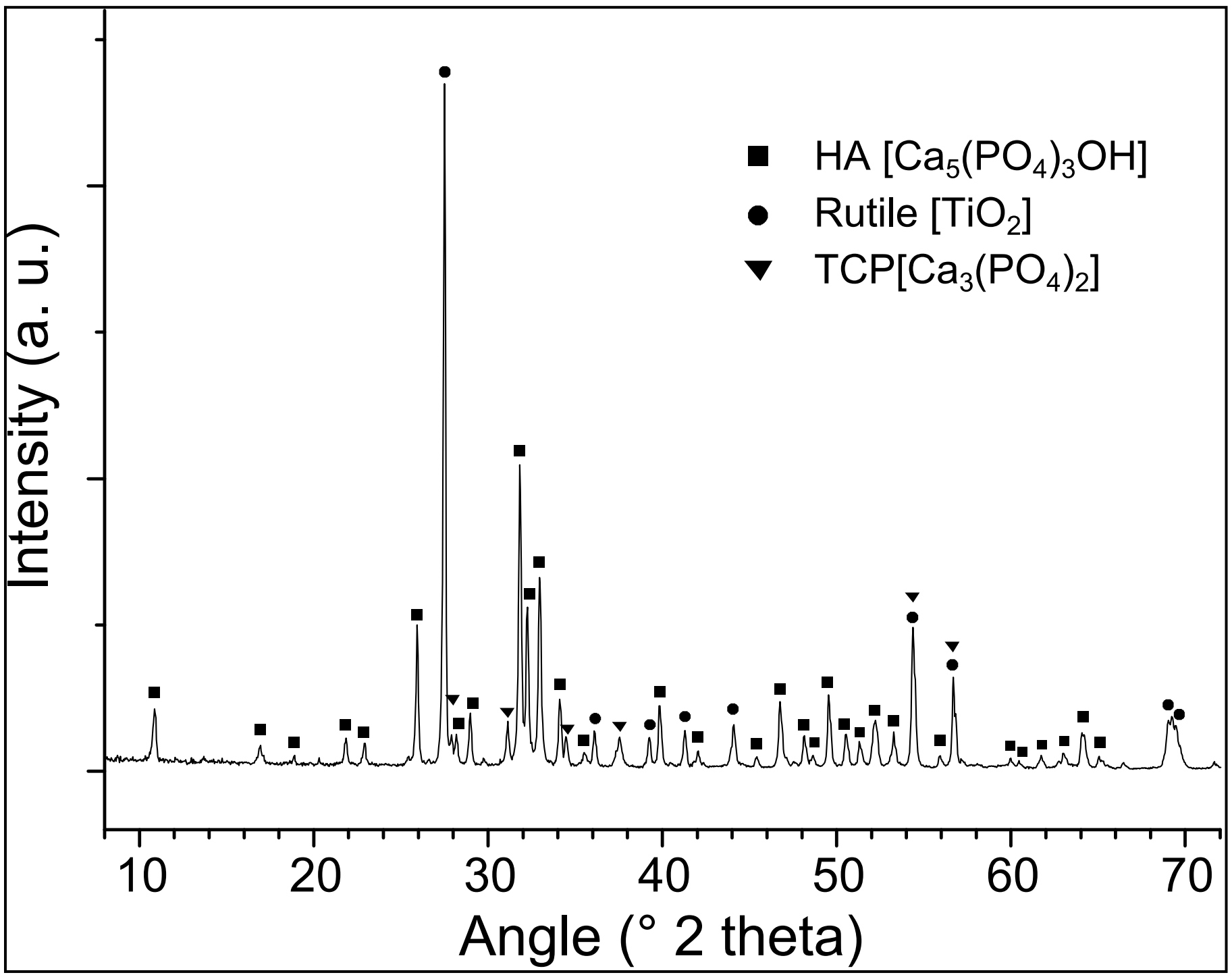

Fig. 1 XRD spectrum of sintered HA coating on Ti plate

## 3.1. SEM analyses

### 3.1.1. DC-EPD

In order to produce samples with increasing coating thickness, the voltage was increased from 220 V to 340 V. The time of deposition was 10 sec in all cases (table 2). The deposition started only from 220 V which could be due to the relatively lower electrophoretic mobility of the particles. Figure 2 presents the evolution of surface (a, b, d, e, g, h, j, k) and cross-section (c, f, i, l) morphologies of DC-EPD coatings with increasing voltage. Cracking was observed in all samples after sintering, which can be attributed to the difference in thermal expansion coefficients of the materials in contact.[15, 18, 21] The cross-section micrographs show that the coating near the Ti substrate is composed of smaller sized particles, which are well packed leading to a denser coating. Further away from the interface, the coating is formed by larger HA particles and becomes more porous.

DC-EPD-1

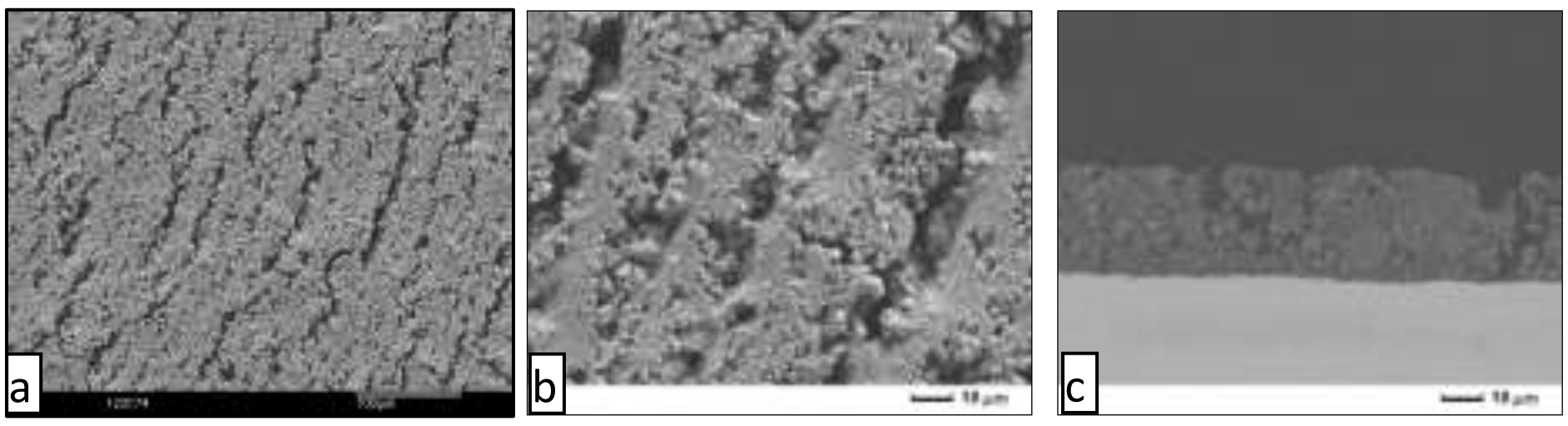

DC-EPD-2

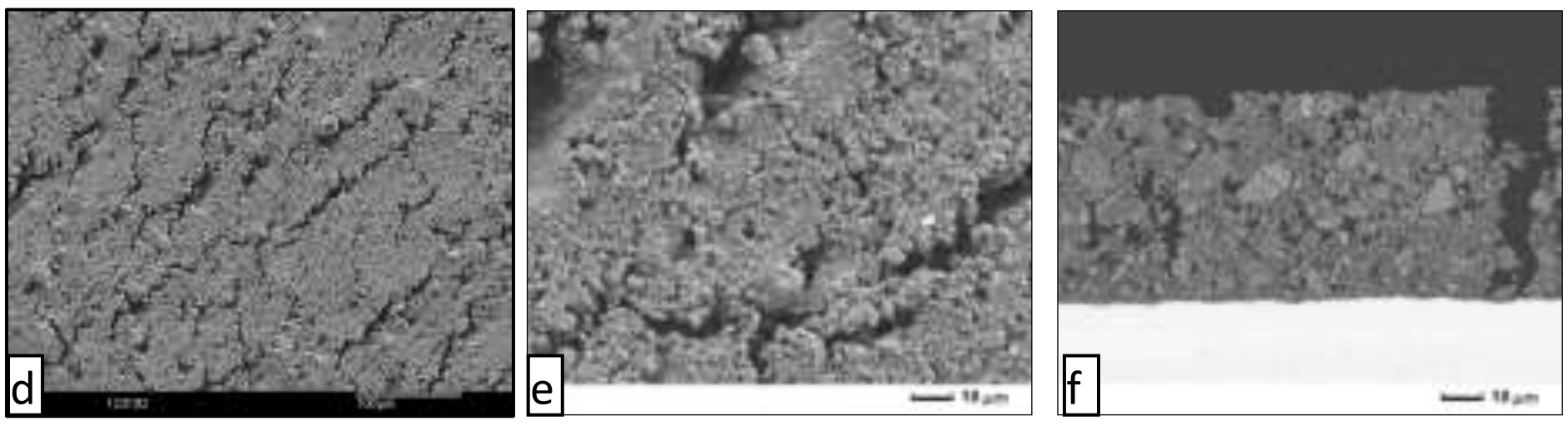

DC-EPD-3

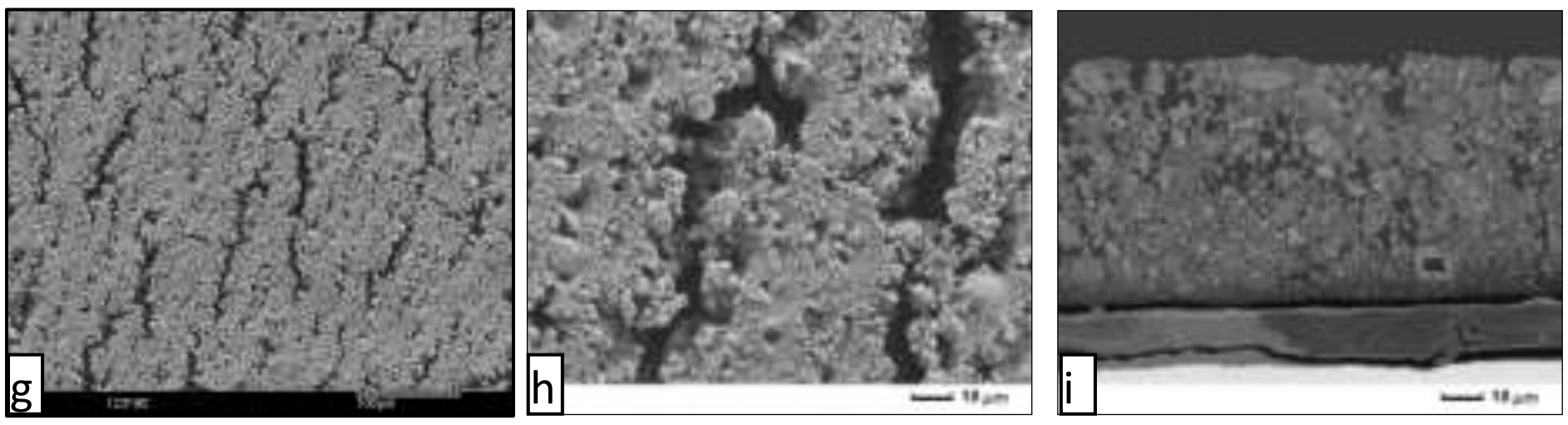

DC-EPD-4

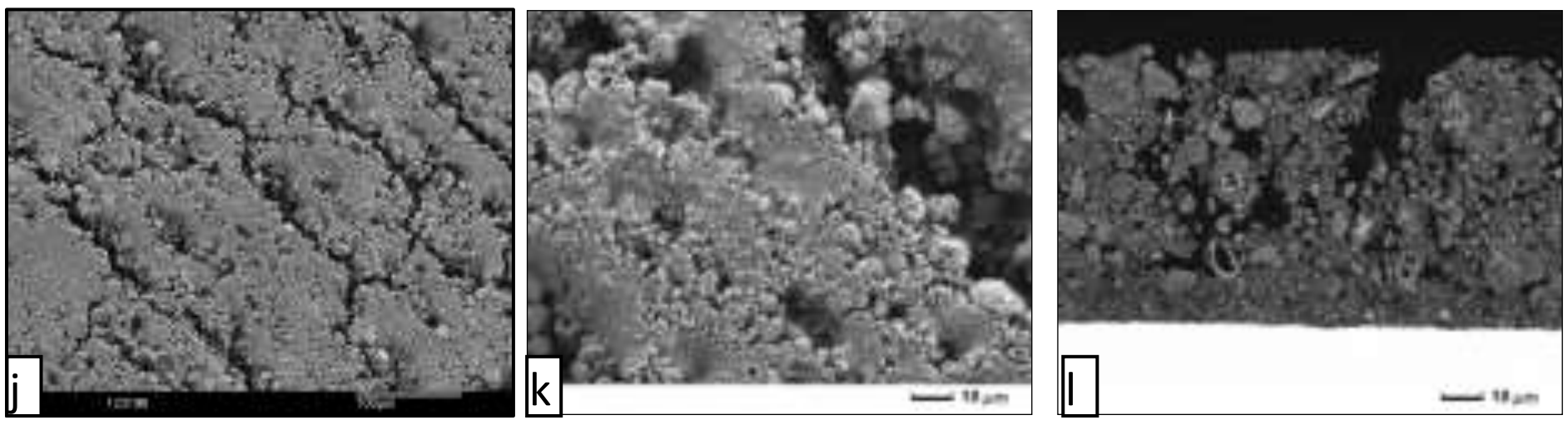

Fig. 2 Scanning electron micrographs of surface and cross-section morphologies of different DC-EPD coatings. The scale bars are (a, d, g, j) 100 µm and (b, c, e, f, g, i, k, l) 10 µm respectively

The cross-section micrographs show that the coating thickness increased with an increase in applied voltage. These thickness values are shown in table 2. The minimum voltage used (220 V) for deposition resulted in a HA coating thickness of 25 ± 2 μm. An effective voltage increase of 40 V almost doubled the thickness. The rate of deposition reduced further with an increase in applied voltage.

An area percentage calculation was performed on surface micrographs in order to understand the evolution of surface crack density. These results are shown in table 2. The values show that the percentages of cracks observed on the DC-EPD samples are similar irrespective of the variation in deposition parameters.

Table 2 Deposition parameters, thicknesses and calculated crack densities of DC-EPD samples

| Sample name | Voltage (V) | Time of coating (sec) | Avg. thickness (μm) | Std. dev. (μm) | Avg. crack area (%) | Std. dev. |
|---|---|---|---|---|---|---|
| DC-EPD-1 | 220 | 10 | 25 | 2 | 36 | 4 |
| DC-EPD-2 | 260 | 10 | 49 | 2 | 34 | 4 |
| DC-EPD-3 | 300 | 10 | 59 | 3 | 34 | 5 |
| DC-EPD-4 | 340 | 10 | 65 | 3 | 32 | 6 |

### 3.1.2. AC-EPD

In order to produce samples with a similar coating thickness to that of DC-EPD coated samples, three different voltage-time combinations were applied. These parameters are shown in table 3. Figure 3 shows the scanning electron micrographs of sintered coatings prepared by AC-EPD technique. The surface morphologies (a, b, d, e, g, h) show a different cracking behavior as compared to the DC-EPD samples. For AC-EPD-1, cracks are minimal as compared to any other samples in this study. However larger cracks were observed more with increasing coating thickness (AC-EPD-2 and AC-EPD-3).

AC-EPD-1

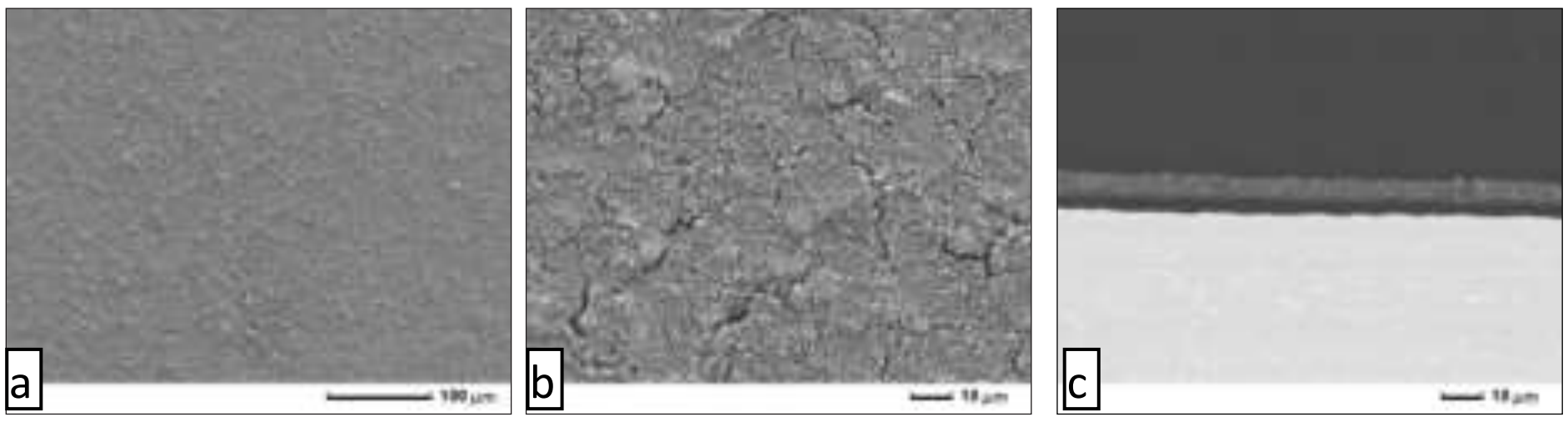

AC-EPD-2

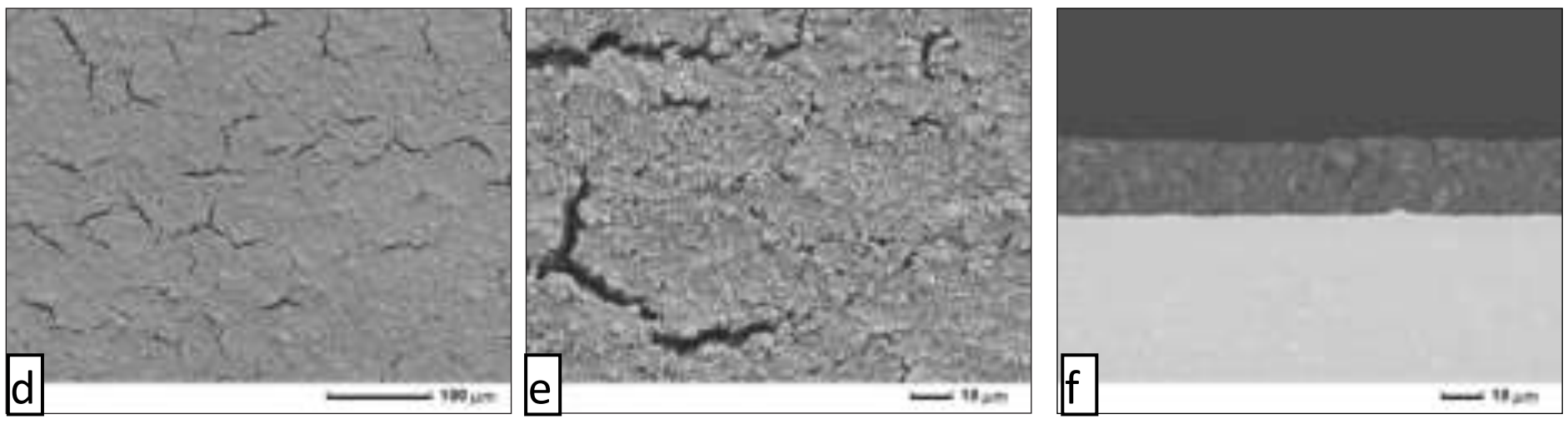

AC-EPD-3

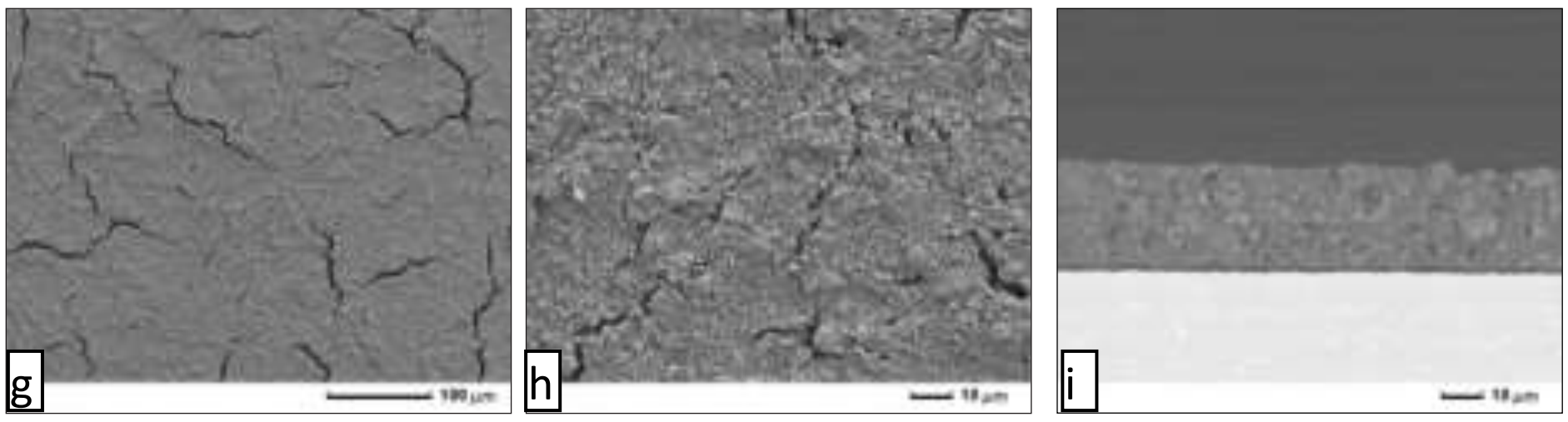

Fig. 3 Scanning electron micrographs of surface and cross-section morphologies of different AC-EPD coatings. The scale bars are (a, d, g) 100 µm and (b, c, e, f, h, i) 10 µm respectively

The thickness values are shown in table 3. There is a sharp increase in thickness of the coatings after an effective voltage increase of 50 V (AC-EPD-2) or time of 30 sec (AC-EPD-3). The area percentages of cracks are also shown in table 3. As in the case of DC-EPD, the deposition parameters in AC-EPD did not change the crack density in a significant manner.

Table 3 Deposition parameters, thicknesses and calculated crack densities of AC-EPD samples

| Sample name | Voltage (V) | Time of coating (sec) | Avg. thickness (µm) | Std. dev. (µm) | Avg. crack area (%) | Std. dev. |
|---|---|---|---|---|---|---|
| AC-EPD-1 | 100 | 30 | 5 | 0.5 | 12 | 3 |
| AC-EPD-2 | 150 | 30 | 20 | 2 | 14 | 2 |
| AC-EPD-3 | 100 | 60 | 22 | 3 | 8 | 2 |

Crack density results shown in table 2 and 3 reveals that the percentage area of surface cracks in case of AC-EPD samples are 20-28 % lower than that of DC-EPD samples. Irrespective of the deposition parameters, the percentage of cracks were similar within the technique used, which indicates that the cracking phenomenon is related to the particle packing and sintering conditions than to the deposition parameters.

### 3.1.3 Porosity calculation

The porosity in the HA coating was measured using image analysis on the cross-section micrographs. To compare the difference in AC and DC coated samples, the samples with similar coating thickness (DC-EPD-1, AC-EPD-3) were selected. The method used was similar to the one described earlier for surface crack area determination. But in this case the ratio between dark and light pixels was calculated after manual thresholding of the selected area. The porosity calculated was 24 % and 11 % respectively for DC-EPD-1 and AC-EPD-3. Thus AC-EPD coated sample is denser which indicates a better packing mechanism in case of AC-EPD method.

## 3.2 Comparison

In order to understand the deposition mechanism, samples with similar thickness (AC-EPD-3 and DC-EPD-1) were selected . The cross-section micrographs show a more densely packed coating for AC-EPD-3. The area percentage of surface cracks in case of AC-EPD-3 was 8 ± 2 % whereas this value was 36 ± 4 % for DC-EPD-1. These cracks are commonly attributed to the difference in thermal expansion coefficients of the materials in contact which leads to thermal stress during the sintering cycle.[15, 18, 22] The difference in thermal expansion coefficients results in different expansion and shrinkage rates of coating and substrate during the firing and cooling steps of sintering. This leads to thermal stress especially at the interface between the coating and substrate. The rutile phase created during the sintering process could also contribute to the thermal stress due to its lower thermal expansion coefficient as compared to HA.[23] Additionally, cracking mechanism could be supported by irregular packing of coating during deposition.

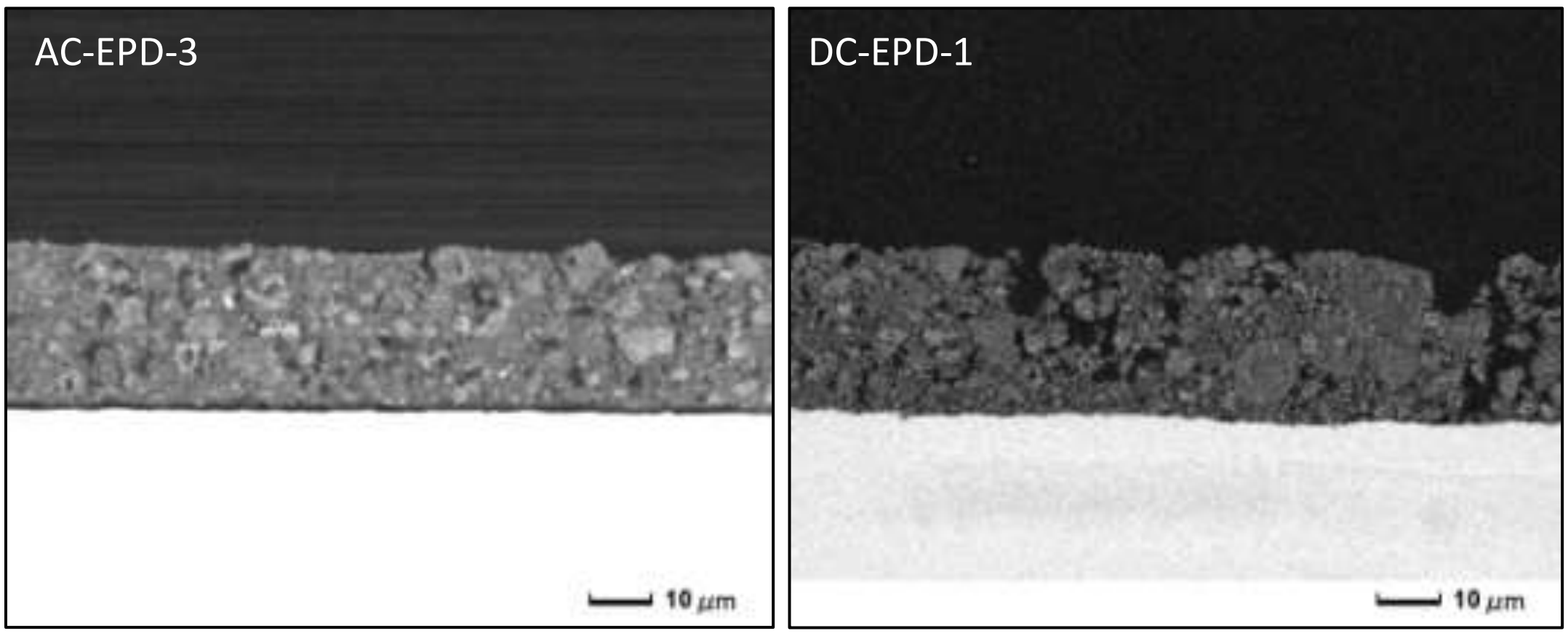

Fig. 4 Comparison of coating cross sections obtained by AC and DC methods

The commercial powder used in this study has a relatively broad particle size distribution (figure A.2). The cross-section image of AC-EPD-3 (figure 4) shows a densely packed coating which is mainly composed of small and medium sized particles. On the other hand more particles of larger size were visible in the case of DC-EPD-1 cross-section (figure 4). The presence of larger particles was more notable in case of thicker coatings obtained via DC-EPD (figure 2l). This observation indicates the possible mechanism happening in either case. During AC-EPD, the particles are experiencing an oscillation according to the applied signal (illustrated in figure 5a). The asymmetry of the wave causes the migration of particles to the Ti substrate. It was observed that applying a symmetric AC signal to the suspension did not result in a deposition, which has also been previously reported.[13] Thus the asymmetry value is important in case of AC-EPD. At relatively high frequencies, the migration probability of smaller particles are higher and hence a better control of the particle size to be coated is possible. This results in a coating which is composed of small and medium sized particles because the migration range of larger particles will be restricted according to the frequency of signal. Due to the oscillation-migration mechanism, there is also a possibility of rearrangement within the deposited particles. This will create a better packing of the particles that leads to a denser coating. For DC-EPD, the particle movement is controlled by a constant electric field. In this case also the smaller particles migrate faster than the larger particles. Since there is no oscillation occurring, all particles could reach the substrate creating a gradient of smaller to larger particle size (figure 5b). The shrinkage of the larger particles deposited on the outer limits of the coating will result in cracking between the loosely bound agglomerates. This type of cracking might be the reason for higher porosity observed in case of DC-EPD coated sample. Either techniques might produce denser and crack free coatings if a powder with narrow particle size distribution is used. Otherwise a multiple coating technique may give rise to more homogeneous coatings as recently reported by Qiang et al. [24] However the advantage of AC-EPD is in producing relatively thick (up to 25 µm in this study) coatings with higher density and lower number of cracks from a powder of broad particle size distribution.

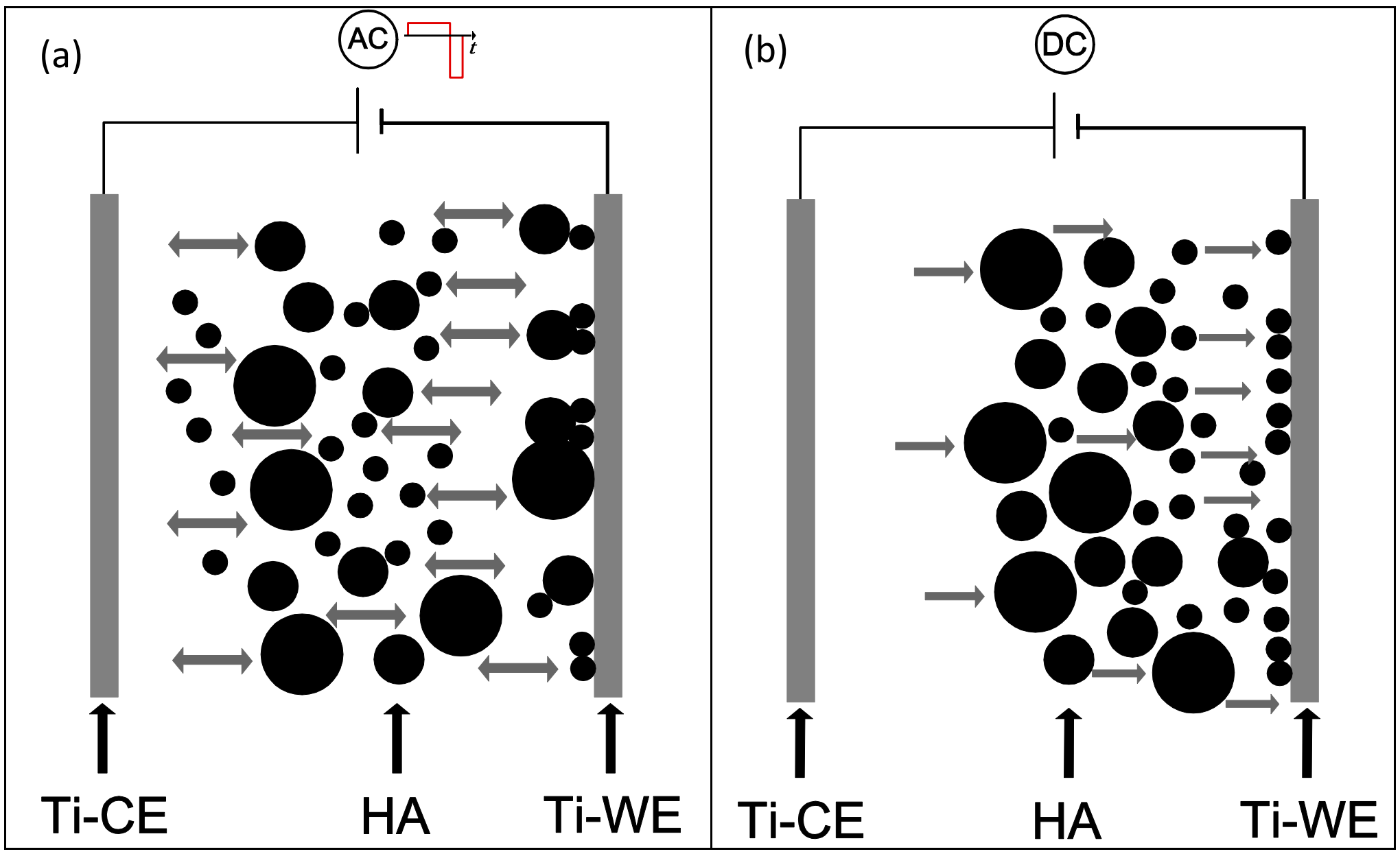

Fig. 5 Suggested particle movement pattern during EPD from (a) AC and (b) DC fields (CE – Counter electrode, WE – Working electrode). An oscillating migration is predicted under AC field resulting in a better packing where as DC field results in a gradient of smaller to larger particle size. See text in § 3.2 for more details

## 3.3. Conclusions

This study compared EPD carried out under AC and DC field conditions. The results showed that AC-EPD method leads to denser and less cracked coatings as compared to DC-EPD at similar thickness. AC-EPD is advantageous for depositing powders with broad particle size distribution as this technique can control the particle migration according to the wave asymmetry and frequency. Fine tuning the asymmetry of AC wave as well as the frequency can further improve the coating characteristics, which will be included in our subsequent study. A powder selection with narrower size distribution profile might increase the packing density of the coatings and the advantage of AC-EPD in such a system is also an interesting case study for future investigations.

## Acknowledgements

The authors gratefully acknowledge the technical assistance of I. Thijs, M. Mertens, M. Schoeters, D. Vanhoyweghen and R. Kemps. VOK wish to thank VITO and ULg for financial assistance, and M. Sharma for proofreading the manuscript.

## References

[1] Sarkar P, Nicholson PS. Electrophoretic deposition (EPD): Mechanisms, kinetics, and applications to ceramics. J Am Ceram Soc 1996; 79:1987-2002.

[2] Van der Biest OO, Vandeperre LJ. Electrophoretic deposition of materials. Annu Rev Mater Sci 1999; 29:327-52.

[3] Boccaccini AR, Zhitomirsky I. Application of electrophoretic and electrolytic deposition techniques in ceramics processing. Curr Opin Solid State Mater Sci 2002; 6:251-60.

[4] Corni I, Ryan MP, Boccaccini AR. Electrophoretic deposition: From traditional ceramics to nanotechnology. J Eur Ceram Soc 2008; 28:1353-67.

[5] Boccaccini AR, Keim S, Ma R, Li Y, Zhitomirsky I. Electrophoretic deposition of biomaterials. J R Soc Interface 2010; 7:S581-613.

[6] Narayanan R, Seshadri SK, Kwon TY, Kim KH. Calcium Phosphate-Based Coatings on Titanium and Its Alloys. J Biomed Mater Res Part B: Appl Biomater 2008; 85B:279-99.

[7] Cadosch D, Chan E, Gautschi OP, Filgueira L. Metal is not inert: Role of metal ions released by biocorrosion in aseptic loosening—Current concepts. J Biomed Mater Res Part A 2009; 91A:1252-62.

[8]. Addison O, Davenport AJ, Newport RJ, Kalra S, Monir M, Mosselmans JFW, Proops D, Martin RA. Do 'passive' medical titanium surfaces deteriorate in service in the absence of wear? J R Soc Interface 2012; 9:3161-64.

[9] Combes C, Rey C. Amorphous calcium phosphates: Synthesis, properties and uses in biomaterials. Acta Biomater 2010; 6:3362-78.

[10] Chávez-Valdez A, Herrmann M, Boccaccini AR. Alternating current electrophoretic deposition (EPD) of TiO2 nanoparticles in aqueous suspensions. J Colloids Interface Sci 2012; 375:102-5.

[11] Chávez-Valdez A, Boccaccini AR. Innovations in electrophoretic deposition: Alternating current and pulsed direct current methods. Electrochim. Acta 2012; 65:70-89.

[12] Trau M, Saville DA, Aksay IA. Field-induced layering of colloidal crystals. Science 1996; 272:706-9.

[13] Neirinck B, Fransaer J, Van der Biest O, Vleugels J. Aqueous electrophoretic deposition in asymmetric AC electric fields (AC–EPD). Electrochem Comm 2009; 11:57-60.

[14] Zhitomirsky I, Gal-or L. Electrophoretic deposition of hydroxyapatite. J Mater Sci: Mater Med 1997; 8:213-19.

[15] Wei M, Ruys AJ, Milthorpe BK, Sorrell CC, Evans JH. Electrophoretic deposition of hydroxyapatite coatings on metal substrates: A nanoparticulate dual-coating approach. J Sol-Gel Sci Tech 2001; 21-39-48.

[16] Drevet R, Fauré J, Benhayoune H. Thermal treatment optimization of electrodeposited hydroxyapatite coatings on Ti6Al4V substrate. Adv Eng Mater 2012; 14:377-82.

[17] Sahoo PK, Soltani S, Wong KC, Chen YC. A survey of thresholding techniques. Comput Vis Graph Image Proc 1988; 41:233-60

[18] Plesingerová B, Súcik G, Maryska M, Horkavcova D. Hydroxyapatite coatings deposited from alcohol suspensions by electrophoretic deposition on titanium substrate. Ceram Silikáty 2007; 51:15-23.

[19] Javidi M, Javadpour S, Bahrololoom ME, Ma J. Electrophoretic deposition of natural hydroxyapatite on medical grade 316L stainless steel. Mater Sci Eng C 2008; 28:1509-15.

[20] Sridhar TM, Kamachi Mudali U, Subbaiyan M. Sintering atmosphere and temperature effects on hydroxyapatite coated type 316L stainless steel. Corr Sci 2003; 45:2337-59.

[21] Mohan L, Durgalakshmi D, Geetha M, Sankara Narayanan TSN, Asokamani R. Electrophoretic deposition of nanocomposite (HAp + $TiO_2$)on titanium alloy for biomedical applications. Ceram Int 2012; 38:3435-43.

[22] Ducheyne P, Van Raemdonck W, Heughebaert JC, Heughebaert M. Structural analysis of hydroxyapatite coatings on titanium. Biomater 1986; 7:97-103.

[23] Meagher EP, Lager GA. Polyhedral thermal expansion in the $TiO_2$ polymorphs: Refinement of the crystal structures of rutile and brookite at high temperature. Can Mineral 1979; 17:77-85.

[24] Chen Q, Cordero-Arias L, Roether JA, Cabanas-Polo S, Virtanen S, Boccaccini AR, Alginate/Bioglass® composite coatings on stainless steel deposited by direct current and alternating current electrophoretic deposition. Surf Coat Tech; doi:10.1016/j.surfcoat.2013.01.042 (in press).

# Appendix A

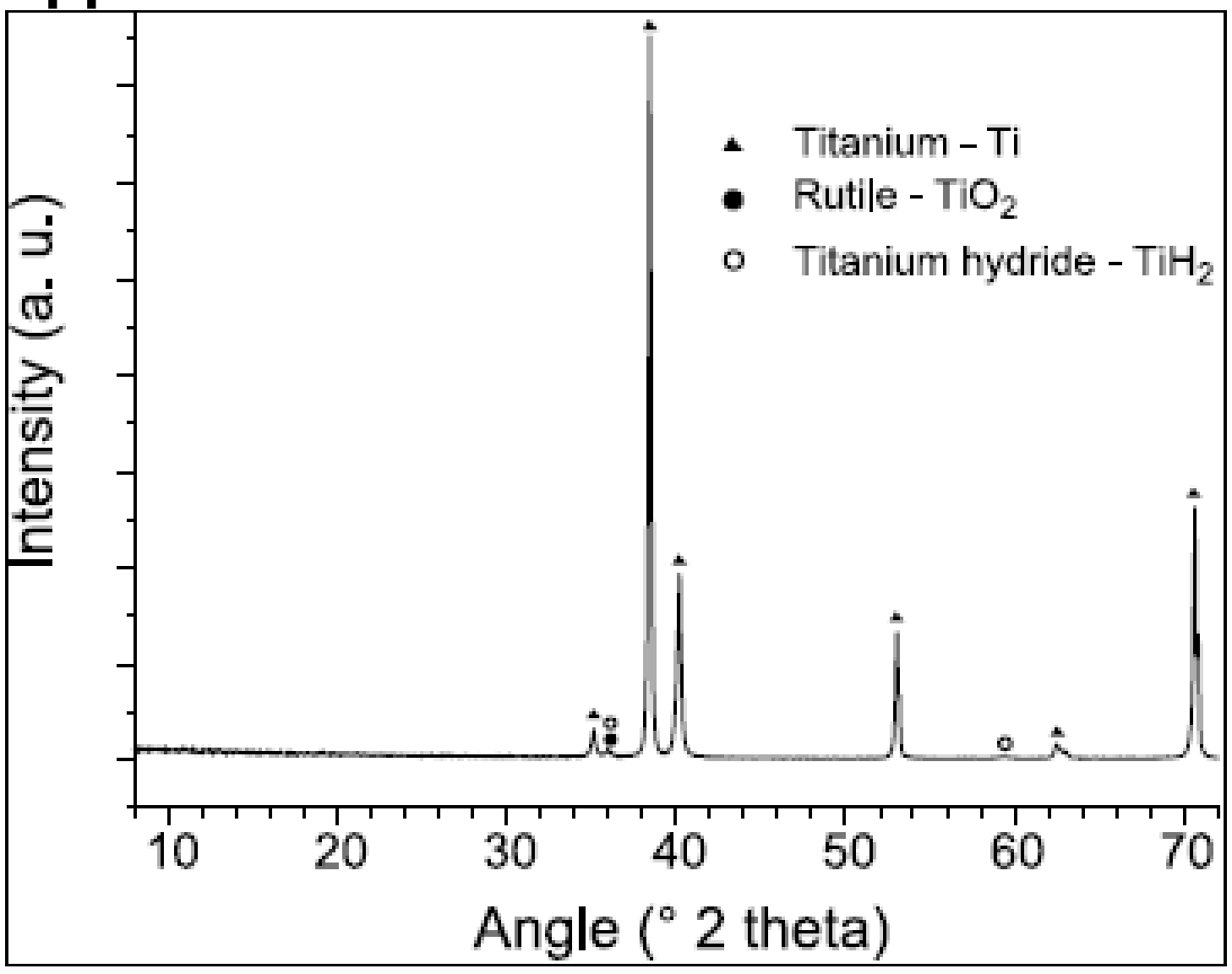

Fig. A1 XRD spectrum of Ti substrate

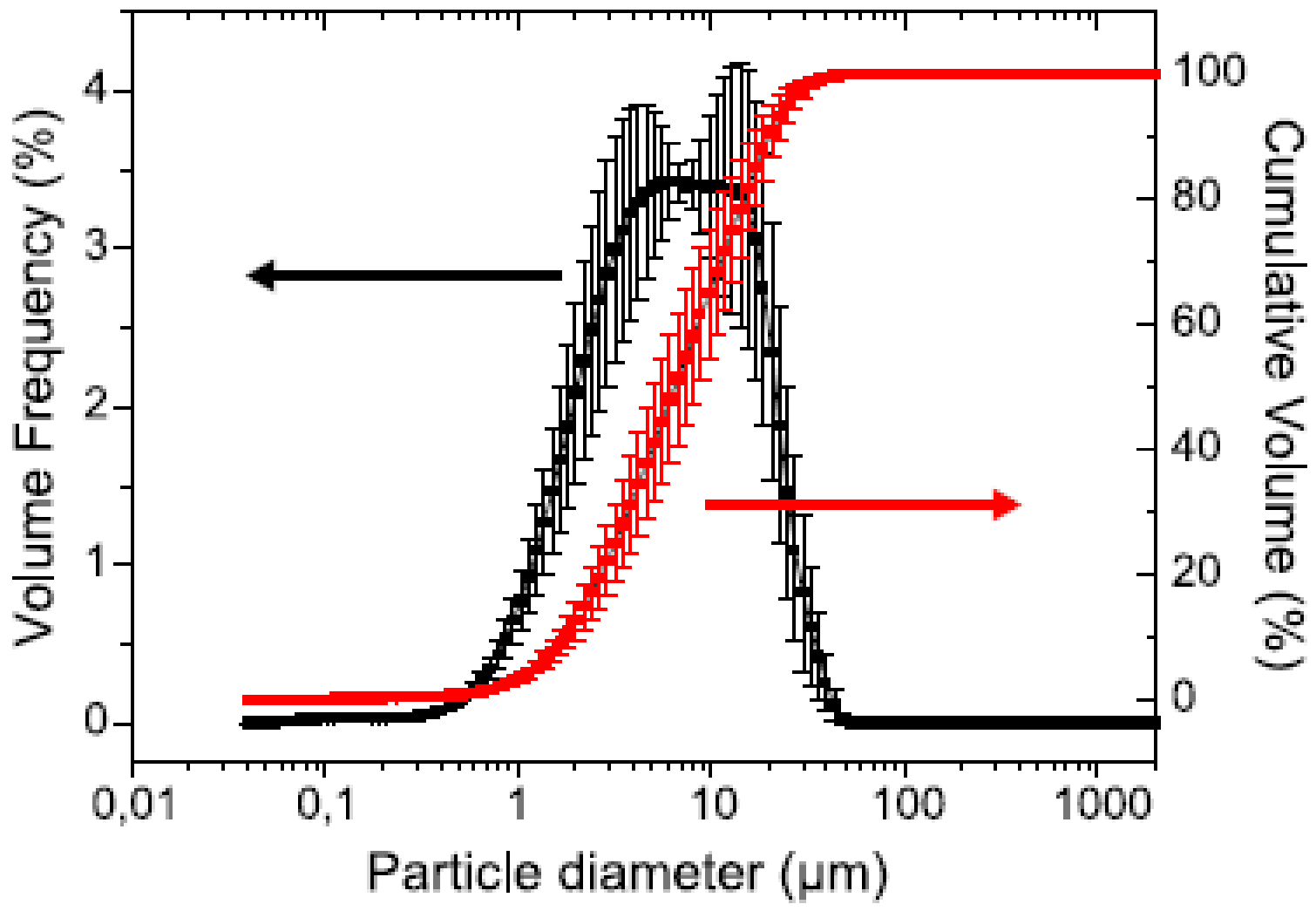

Fig. A2 Particle size distribution of HA used in this study